\documentclass[twocolumn,runningheads]{svjour2}
\smartqed  
\usepackage{graphicx}  
\usepackage{latexsym,amssymb}

\newcommand {\ergcms} {erg\,cm$^{-2}$s$^{-1}~$}

\newcommand {\hess} {H.E.S.S. }

\newcommand {\m} {$\;\rm \mu m$}

\newcommand {\nw} {nW m$^{-2}$ sr$^{-1}$}

\journalname{Astrophysics and Space Science}
\begin{document}

\title{A low density of the Extragalactic Background Light
revealed by the H.E.S.S. spectra 
of the BLLac objects 1ES\,1101-232 and H\,2356-309}


\titlerunning{EBL limits from H.E.S.S. BLLac spectra}        

\author{Luigi Costamante (H.E.S.S. collaboration)
}


\institute{L. Costamante \at
              Max-Planck-Institut f\"ur Kernphysik \\
	      Saupfercheckweg 1, D-69117 Heidelberg, Germany \\
              Tel.: +49-6221-516470\\
              \email{luigi.costamante@mpi-hd.mpg.de}           
}

\date{Received: date / Accepted: date}

\maketitle

\vspace*{-0.5cm}
\begin{abstract}
The study of the TeV emission from extragalactic sources is hindered 
by the uncertainties on the diffuse Extragalactic Background Light (EBL). 
The recent H.E.S.S. results on the blazars 1ES\,1101-232 and H\,2356-309 represent
a breakthrough on this issue.  Their unexpectedly hard spectra allow an upper 
limit to be derived on the EBL in the optical/near-infrared range,  
which is very close to the lower limit given by the resolved galaxy counts. 
This result seems to exclude a large contribution to the EBL from other sources 
(e.g. Population III stars) and indicates that the intergalactic space 
is more transparent to $\gamma$-rays than previously thought. 
A discussion of EBL absorption effects
and further observational tests with Cherenkov telescopes are presented.
\keywords{$\gamma$-rays \and AGN \and EBL \and extragalactic \and diffuse background \and blazar}
\PACS{98.54.Cm \and  98.62.Ai  \and 95.85.Kr  \and 95.85.Jq \and 95.85.Pw}
\end{abstract}

\begin{figure}
\vspace*{-0.5cm}
\centering
 \includegraphics*[width=0.4\textwidth]{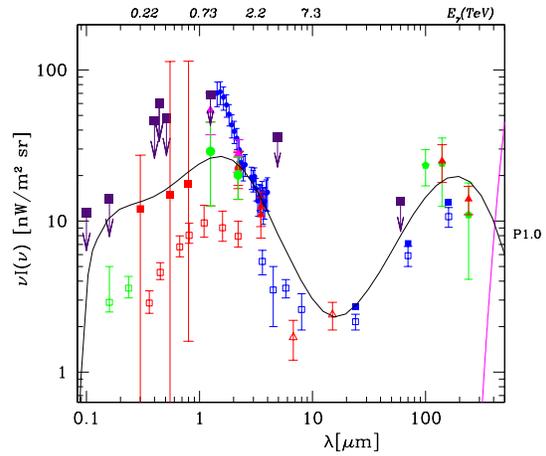}
\caption{Spectral Energy Distribution of the diffuse Extragalactic Background Light.
Open symbols correspond correspond to the integrated light from resolved galaxies,
and thus represent lower limits for the EBL (refs in \cite{nature}).
From left to right: HST data (green and red squares),
Spitzer data (blue squares), ISO data (red triangles\cite{elbaz}),
Spitzer data (blue squares\cite{dole}; small filled squares are the corresponding 
estimates for the total galaxy emission). 
The other filled symbols show the different EBL estimates
(compilation from \cite{hauser,nature,mattila}).
Above 300\m\ the CMB starts to dominate. The black line shows the double-humped 
phenomenological curve used as reference shape for the EBL SED (see \cite{icrc,nature}).
The upper axis shows the energy of the $\gamma$-ray photon which interact with the corresponding 
EBL photon at the peak of the $\gamma\gamma$ cross-section. 
}
\label{eblsed}       
\vspace*{-0.4cm}
\end{figure}

\vspace*{-0.5cm}
\section{Introduction}
\label{intro}
The diffuse Extragalactic Background Light (EBL) contains
unique information about the epochs of formation
and the history of evolution of galaxies, since it consists 
of the sum of the light produced by all extragalactic sources over 
cosmic time. Its Spectral Energy Distribution (SED, Fig. \ref{eblsed}) is 
characterized by two broad bumps\cite{hauser}, produced by thermal emission contributed 
directly from stars at optical (Opt) and near infrared (NIR) wavelengths,
and by dust which partly absorbs and re-emits the starlight 
at longer wavelengths (FIR).
The Opt-NIR range 
has gained special interest in recent years because 
some direct measurements between 1 and 3\m\
have suggested the presence of a significant excess 
above the contribution
of normal galaxies, which has been intriguingly interpreted 
as the signature of redshifted UV light from heavy 
Population-III star formation at $z\sim10$ \cite{santos} 
(though with severe energy-budget problems, see \cite{madausilk}).
Unfortunately, direct EBL measurements are subject to
possible large systematic uncertainties due to the
difficulty of an accurate modelling and then subtraction of the
bright foregrounds, mainly consisting of zodiacal (interplanetary dust)
light\cite{hauser}.

\begin{figure*}
\vspace*{-0.6cm}
\centering
  \includegraphics[width=0.34\textwidth]{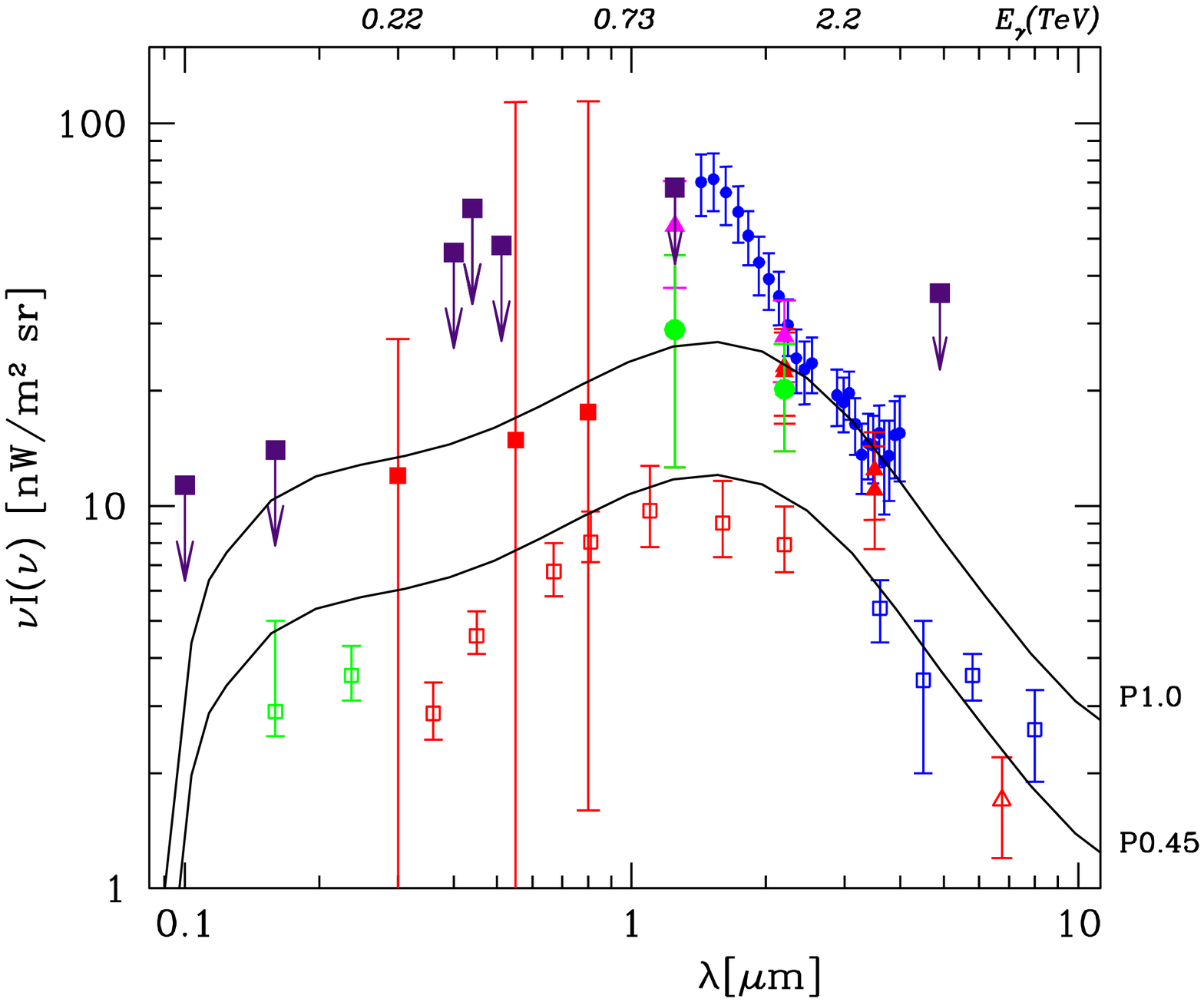}
  \includegraphics[width=0.31\textwidth]{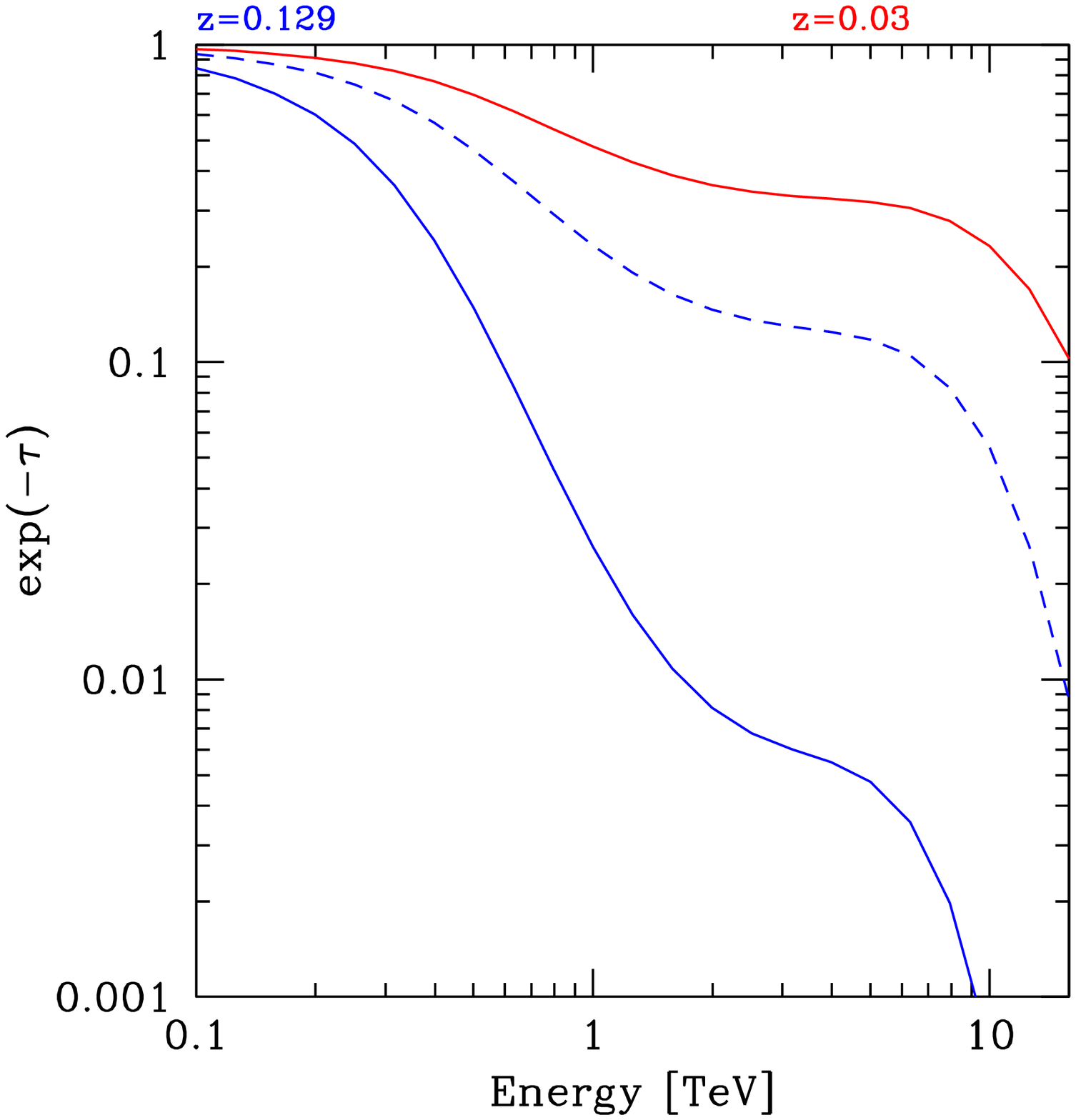}
  \includegraphics[width=0.31\textwidth]{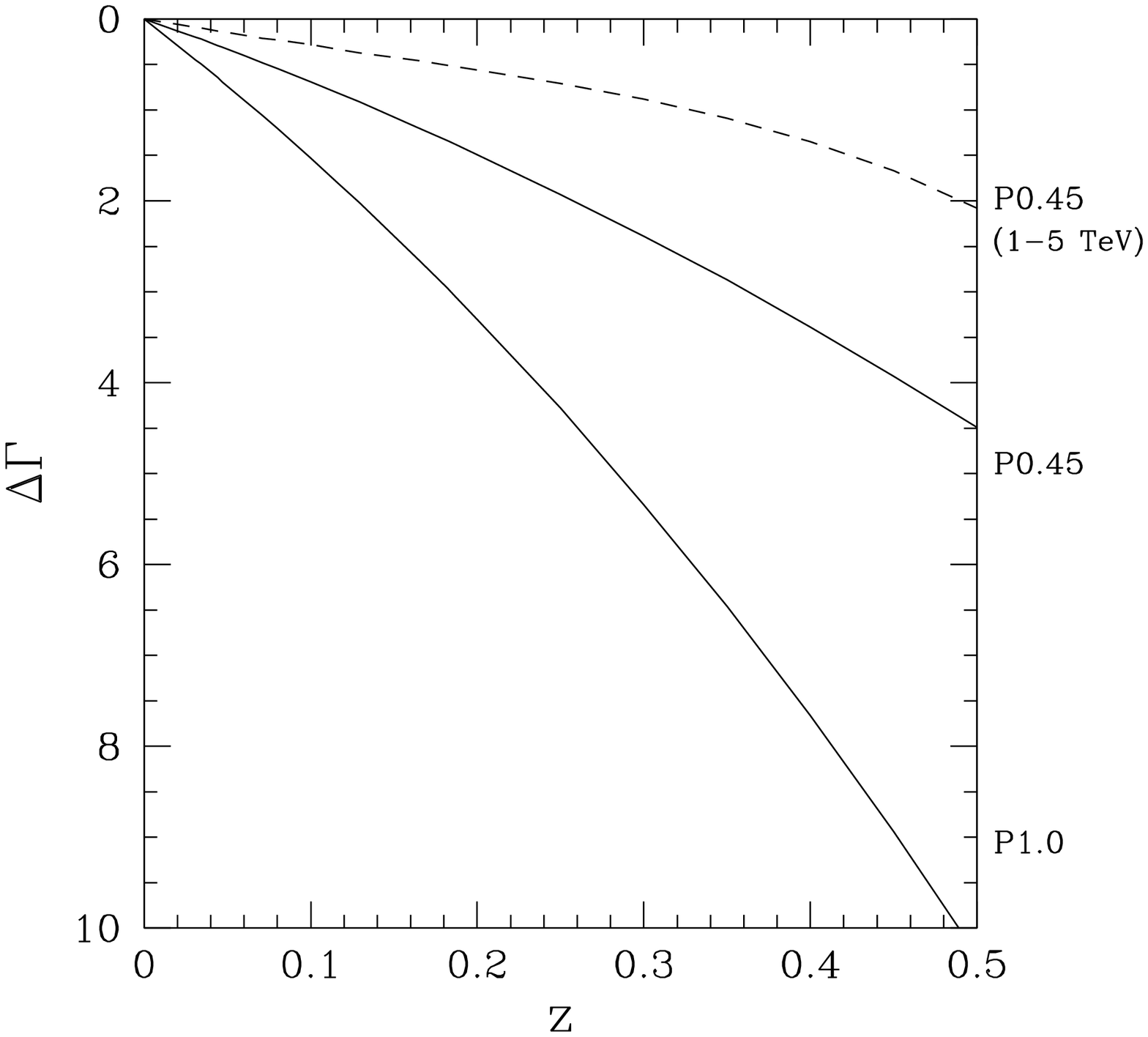}
\vspace*{-0.3cm}
\caption{Effects of changing the EBL normalization
(expressed as scaling factor P: left panel, 
1.0$\times$ and 0.45$\times$) 
on the attenuation factor $e^{-\tau}$,
for a fixed redshift (center panel: full and dashed blue lines, respectively).
In the center panel, the effect of increasing redshift is also shown
(red and blue full lines, respectively) for a fixed EBL SED (P1.0).
The attenuation curves represent directly the shape of an observed VHE spectrum, if the source 
has an emission spectrum parallel to the upper axis. All calculations were performed assuming
$H_0=70\;\rm km/s/Mpc$, $\Omega_m=0.3$, $\Omega_{\Lambda}=0.7$.
Right panel (full lines): the difference 
in photon index between emitted and observed spectrum (when fitted with a power-law model between 
0.2 and 2 TeV) is plotted as a function of redshift, for two EBL normalizations.
The dashed line shows the milder deformation of the original spectrum in the energy range
1-5 TeV, due to the $\propto \lambda^{-1}$ EBL slope between 2-3 and 10\m.
}
\vspace*{-0.2cm}
\label{fig2}       
\end{figure*}

An alternative approach\cite{icrc} is provided by the measurement of very high energy
(VHE) $\gamma$-rays from extragalactic sources,
through the detection and identification of absorption features
in the VHE spectra caused by the interaction with EBL photons 
($\bf\gamma\gamma\rightarrow e^{+}e^{-}$).
High-energy peaked blazars (HBL) represent to this respect excellent sources
of $\gamma$-ray beams, thanks to their prominent TeV emission and high apparent luminosity,
which makes them detectable from large distances and/or with good photon statistics.
The ``downside" 
is that they are highly variable 
and characterized by a very wide range of possible spectra. 
Though a lot has been learned about them, the present understanding of their radiation 
processes is not yet complete enough to reliably predict the intrinsic TeV spectra,
and thus to disentangle absorption from intrinsic features.
Conversely, the still large uncertainties on the EBL 
prevent a sufficiently accurate reconstruction of their intrinsic spectra 
in order to constrain the emission models.
The studies of blazars and EBL are therefore tightly coupled,
in a classic case of ``one equation, two variables" problem\cite{icrc,costa}.

So far, only hints and guesses could be made
from the TeV detected objects (in particular from the more distant 
1ES 1426+428 \cite{hegra} and PKS 2155-304 \cite{hess2155,dwek}), since the reconstructed spectra
could be generally explained also as blazars' intrinsic features.
The surprising hardness of the \hess spectra of 1ES\,1101-232 
and H\,2356-309\cite{nature,matthias}) represent a breakthrough in this respect,
posing the problem of conflict between blazar spectra and EBL estimates
in a much more severe and compelling way.

\begin{figure*}
\vspace*{-0.6cm}
\centering
  \includegraphics[width=0.34\textwidth]{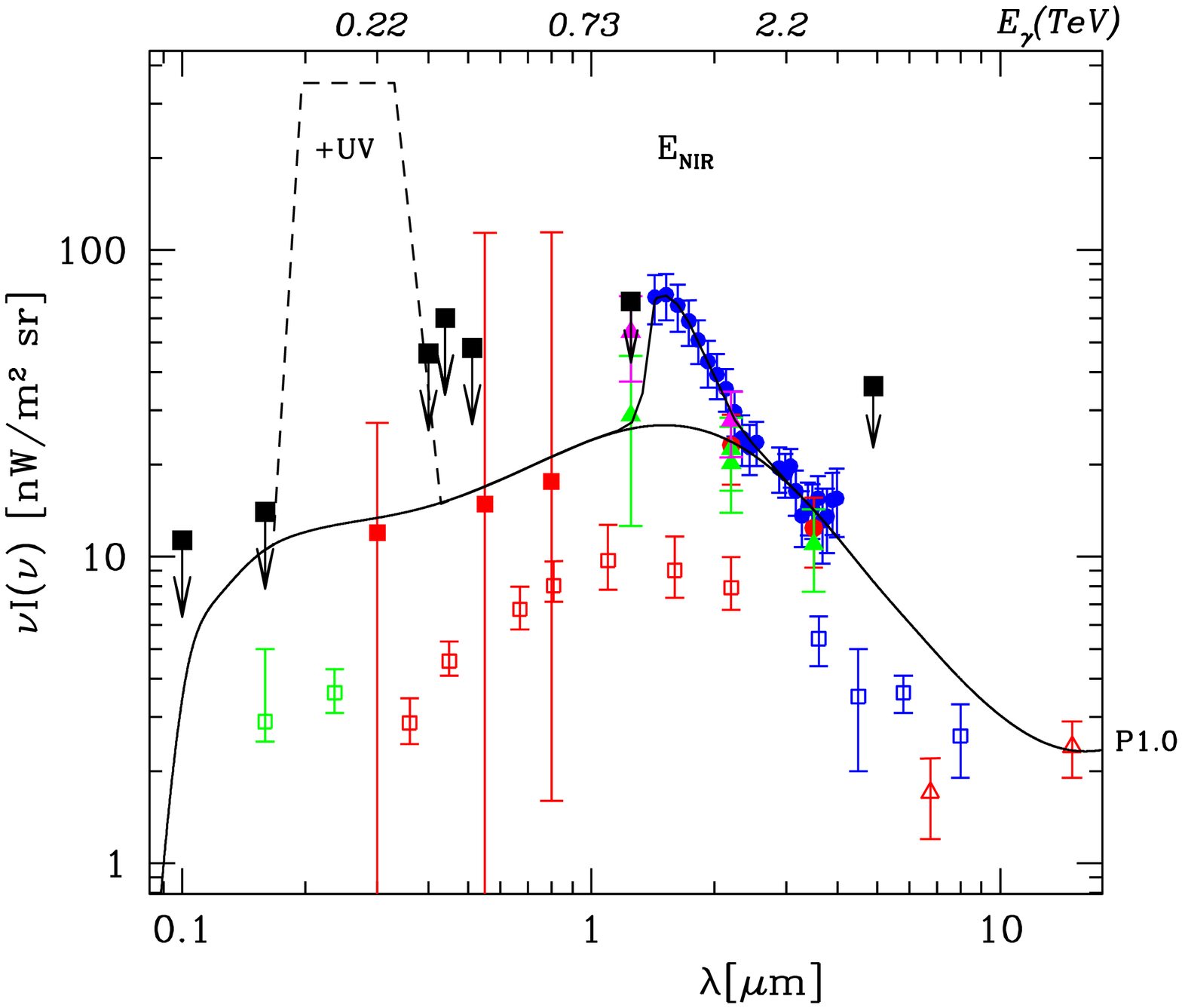}
  \includegraphics[width=0.31\textwidth]{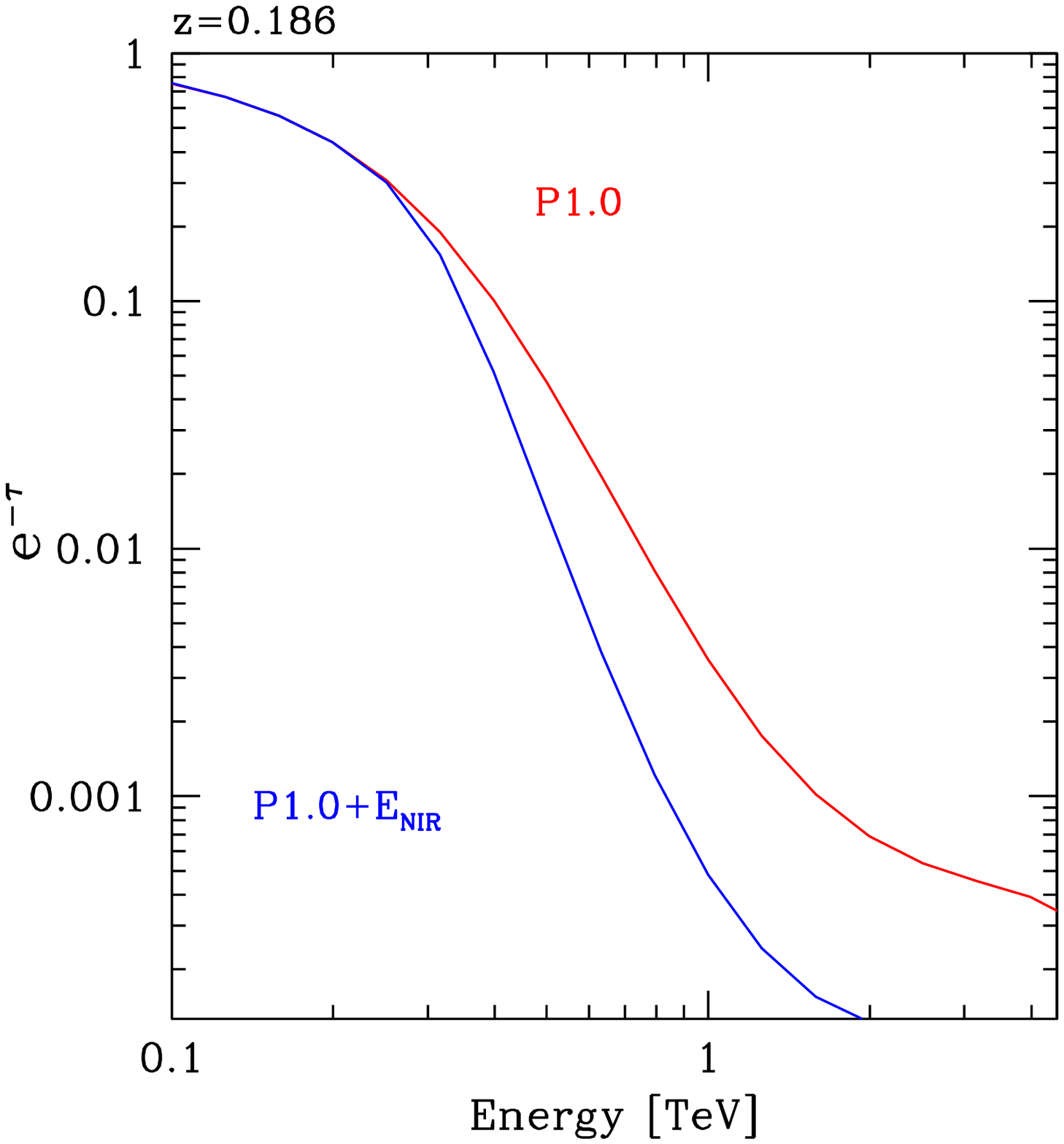}
  \includegraphics[width=0.31\textwidth]{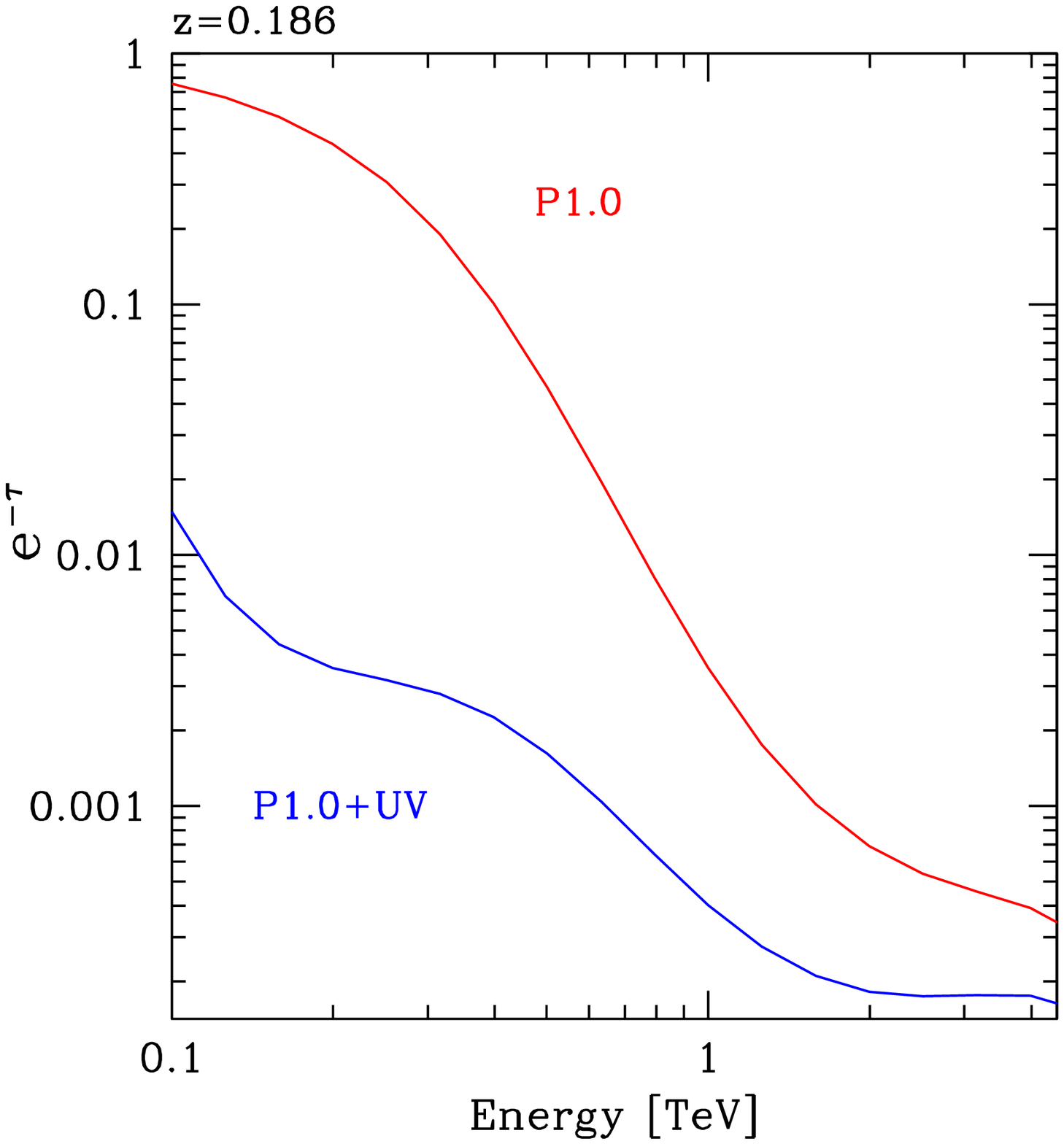}
\vspace*{-0.3cm}
\caption{Effects of changing the EBL spectral shape 
(increasing the flux selectively at 1-2\m\ or at 0.2-0.3\m, left panel) 
on the attenuation factor, for fixed redshift. 
Center panel: an increase only at 1-2\m\ strongly steepens the spectrum, since it increases
the optical depth at $\sim1-2$ TeV without affecting the 0.2-0.3 TeV range
(since below threshold).
Instead, a higher EBL flux in the UV range alone increases absorption at 0.2 TeV
more than at 1-2 TeV, thus reducing the energy dependence of the optical depth
(right panel). As a result, the spectrum suffers less steepening, 
at the price of an overall higher attenuation.
}
\vspace*{-0.2cm}
\label{fig3}       
\end{figure*}

\section{EBL effects on TeV $\gamma$-ray spectra}
\label{sec1}
To illustrate the effects of absorption, 
as reference shape for the EBL SED we adopted  the phenomenological 
curve used in \cite{icrc,hegra} 
(labelled P1.0 in Figure 1),
which is designed to be in general agreement with the EBL spectrum expected 
from galaxy emission\cite{prim01} (modulo normalization).
Such EBL SED imprints a specific shape onto the original spectrum, 
as illustrated in Fig. 2 (center panel): 
above $\sim$0.1 up to $\sim$1-2 TeV it causes a strong steepening,
followed by a flattening between 2 and $\sim$7 TeV, where the observed spectrum partially 
recovers the original slope (owing to the decline approximately $\propto\lambda^{-1}$
above few microns, which causes the optical depth to become nearly 
constant with energy\cite{icrc}). Further on, a sharp cut-off occurs above $\sim$8-10 TeV 
caused by the EBL SED rising again above 10\m.

This yields two consequences: {\it a)} with the current systematic uncertainties 
of Cherenkov telescopes (CT), if the intrinsic spectrum is a power-law
the observed spectrum can still be well approximated by a power-law model
in the range 0.2-2 TeV, but of steeper spectral index ($\Gamma_{obs}>\Gamma_{int}$); 
{\it b)} measuring a spectrum above $\sim$200 GeV, {\it no cut-off} is expected from EBL absorption
up to $\sim$6-7 TeV: any cut-off measured by CT in such energy range is more likely
to be intrinsic (unless considering an EBL NIR spectrum much 
flatter than $\propto\lambda^{-1}$).
Note also that in the range 0.2-10 TeV 
breaks or cut-offs in the emitting particle distribution of  blazar jets
are quite common, as shown by the corresponding synchrotron emission in the Opt--X-ray range.
Such intrinsic steepening might cancel out the expected flattening
due to EBL absorption. In other words, the absence of such flattening in the TeV spectra of
some blazars does {\it not} imply an evidence for a different EBL spectrum.
To detect such feature, a power-law (and possibly hard) intrinsic TeV spectrum 
over $\sim$2 decades is generally needed (e.g. 1ES 1426+428\cite{hegra}).

Figure 2 shows the effect of a change of the EBL normalization ``P" and of source redshift:
$\Delta\Gamma$ (the difference between observed and intrinsic slope)
increases with both normalization and redshift.
The two dependencies combine, so that at higher redshifts
an equal change 
of EBL density yields a much stronger effect (Fig. 2, right panel). 
Redshift thus provides leverage: more distant objects represent therefore
a more sensitive diagnostic tool for EBL studies, but with the limitations 
given by generally lower fluxes (and thus statistics). 

Figure 3 shows instead the effects 
of a change in the EBL {\it spectrum} with respect to our template, 
namely increasing the flux in the 1-3\m\ range
(as would be required to match some  direct estimates, 
and expected in the Pop III stars scenario) or in the 0.1-0.3\m\ range.
In the first case the steepening gets stronger, while in the second case
it is reduced, because of a lower contrast in optical depth
between 0.2 and 1 TeV photons (though at larger absolute values).
Note that a {\it decrease} of the UV-opt fluxes would not reduce 
the steepening, since it increases the contrast between the two energies
while reducing the absolute values.

It is also worth to highlight an 
important consequence of
EBL absorption on blazar interpretation:
if the EBL is high, sources at $z\gtrsim0.1$ would be so heavily absorbed 
that their intrinsic Inverse Compton (IC) luminosity and peak energy 
become always much higher than the observed ones. 
The absorption-corrected values can easily reach IC peak energies $>3$-10 TeV 
and luminosities in excess of $10^{47-48}$\ergcms, with a Compton dominance of 10-100.
Therefore, the bulk of their luminosity would be emitted at high energies, alike FSRQ.
In other words, they would represent the famous ``high-energy peaked, high-luminosity" 
blazars that would invalidate the blazar sequence scenario\cite{gg98,padovani} 
(though it 
would remain the issue of
why their radio luminosity is so low compared to FSRQ).
These objects would not have been recognized as such counter-examples 
simply because their luminosity 
would be largely underestimated, being dispersed in the intergalactic space 
due to $\gamma-\gamma$ absorption.
Though now this seems not the case (see Sect. 3),
this possibility still remains also with a low EBL level, 
considering  objects at significantly larger redshift 
(but which are likely beyond the horizon of the present generation of CT).

\begin{figure}
\centering
  \includegraphics*[width=0.4\textwidth]{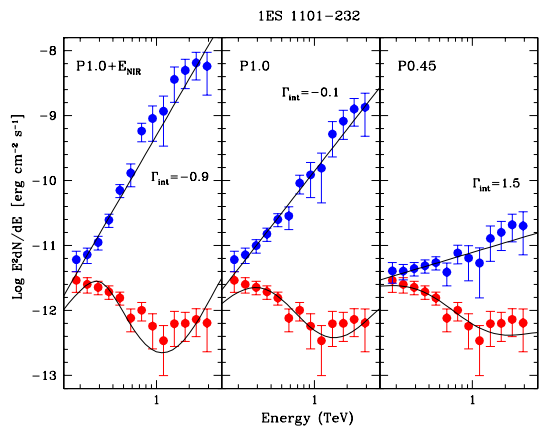}
\caption{The HESS spectra of 1ES\,1101-232, corrected for absorption 
with three different EBL SEDs, as labelled (from \cite{nature}, see ).
Red: observed data. Blue: absorption-corrected data.
The lines show the best fit power-laws to the reconstructed spectrum, 
and the corresponding shape after absorption
}
\vspace*{-0.1cm}
\label{fig4}       
\end{figure}

\section{EBL upper limits from H.E.S.S. spectra}
The intrinsic TeV spectra were reconstructed directly from the observed ones,
using the optical depth calculated for the assumed EBL SED. 
In this way no {\it a priori} assumption on the blazar spectrum
is required. The reconstructed spectra are compatible with a power-law model, 
but high EBL fluxes yield extremely hard photon indices 
($\Gamma_{\rm int}\lesssim0$), for both 1ES\,1101-232 (Fig. 4, and whole SED in Fig. 5)
and H\,2356-309 (see \cite{nature}).
Such hard spectra have never been seen in the closest, less absorbed 
objects (e.g. Mkn~421 and Mkn~501, $\Gamma_{\rm int}\approx1.5-2.8$ 
for the same EBL SEDs) and are difficult to explain  within  
the standard leptonic or hadronic scenarios\cite{icrc} for blazar emission. 
Assuming instead more typical values for the intrinsic spectrum as 
expected from the currently known blazar physics and phenomenology,
namely  that the true average VHE spectrum during the \hess observations
was $\Gamma_{\rm int}\gtrsim1.5$ (as obtained under most circumstances
if the index for the particle spectrum in shock acceleration 
models is $s\geq$1.5, \cite{malkov}), the EBL SED has to be scaled down
to the level of P0.45, very close to the galaxy counts limit (details in \cite{nature}).
This represents the upper limit for the EBL in the Opt-NIR range
which does not require dramatic changes for the blazar physics
or in the scenarios for the underlying particle spectra.
Accounting for galaxy evolution and for the statistical and systematic 
uncertainties on the \hess spectral measurement, the limit corresponds
to $\lesssim(14\pm4)$\,\nw  (i.e. $\leq P0.55\pm0.15$; Fig. 6).
Of course, a different limit can be derived according to 
different hypotheses, and depending on their overall plausibility.
But only strong spectral differences 
would qualitatively change this result (see Sect. 4): 
for example, even the case of a particle spectrum with $s=1$,
which has been argued as asymptotically possible in finit-Mach-number shocks
under certain conditions (see e.g. \cite{schick2,vainio}), 
would raise the limit only slightly
(for 1ES\,1101-232, the change of scaling factor P
is approximately $\Delta{\rm P}\simeq0.34\Delta\Gamma$).
\begin{figure}
\vspace*{-0.2cm}
\centering
  \includegraphics*[width=0.43\textwidth]{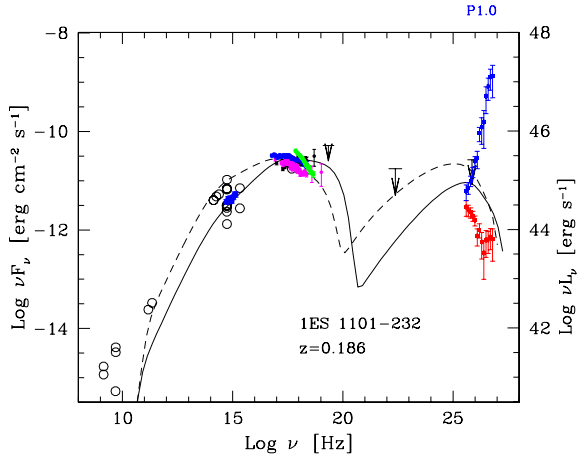}
\caption{SED of 1ES\,1101-232\cite{wolter}, for an EBL spectrum
equal to P1.0. At lower energies, XMM data (blue) and RXTE data (green)
have been taken in the same epoch of the \hess observations (Aharonian et al. 2006, in prep.). 
Magenta: historical SAX observations\cite{wolter}.}
\label{fig5}       
\vspace*{-0.3cm}
\end{figure}
\section{Alternative scenarios}
Another way  to avoid such hard spectra is to reduce 
the energy dependence of the optical depth by increasing
absorption at low $\gamma$-ray energies, i.e. with high UV-Opt fluxes.
However, the huge fluxes required to have $\Gamma=1.5$ 
($>300$ \nw, see \cite{nature} and Fig. 3 left) are in contrast with 
measurements\cite{bernstein}, even when interpreted as upper limits, 
and could hardly be accomodated within any reasonable cosmological model.
Unless considering even more exotic scenarios like
violation of Lorentz invariance or the non-cosmological origin of blazars' redshift,
at present the most viable alternative is that 
such hard spectra are a real, newly discovered feature of the blazar emission.
\begin{figure}
\centering
  \includegraphics*[width=0.4\textwidth]{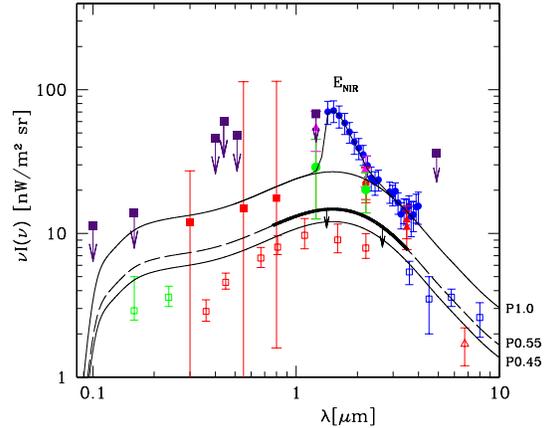}
\caption{Upper limit for the EBL (black marked region), 
corresponding to the assumption of an
intrinsic 1ES\,1101-232 spectrum with photon index $\Gamma \geq 1.5$.
It is obtained from the P0.45 curve, when accounting for galaxy 
evolution effects (details in \cite{nature}).}
\label{fig6}       
\vspace*{-0.2cm}
\end{figure}
Possible mechanisms have already been envisaged\cite{icrc}.
Bulk-motion Comptonization in the deep Klein-Nishina regime
of a narrow-band photon field (such as a Planck-type distribution) by cold plasma
with a very large Lorentz factor may lead to very hard spectra with 
a sharp pile-up, reproducing spectra like the ones in Fig. 4,5 (see \cite{icrc}).
A ``pile-up" (i.e. relativistic Maxwellian) particle distribution seems also expected as
outcome of turbulent acceleration (see e.g. \cite{hs,pp} and refs therein).
In this case, however, even suppressing cooling and considering a purely monoenergetic
distribution (or a distribution with a similarly sharp low energy cut-off at $\gamma_{min}>10^5$ 
\cite{katar}), the resulting synchrotron and SSC spectrum
would be at most $\propto \nu^{1/3}$, i.e. $\Gamma\geq0.66$ (\cite{katar}).
If so, the EBL P0.45 upper limit would be shifted to $\sim$P0.72.
Note that to make this scenario feasible it is necessary 
to always hide the hard synchrotron emission of these particles
below a second, broader component which dominates the blazar emission at low energies,
in order to account for the simultaneous SED (Fig. 5).
\begin{figure}
\centering
  \includegraphics*[width=0.35\textwidth]{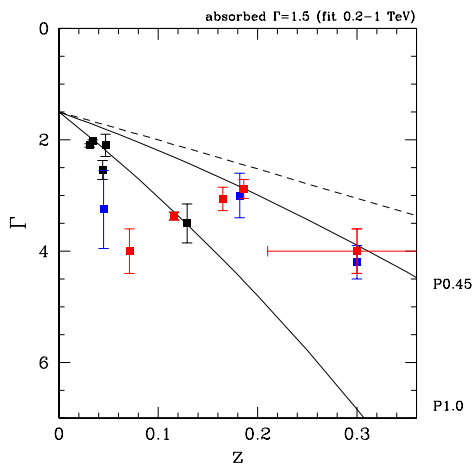}
\caption{Observed hardest photon indices of the VHE spectra of all detected TeV blazars 
as a function of redshift. The lines give the expected observed (i.e. absorbed) photon index
of a source with an intrinsic spectrum $\Gamma=1.5$,
when fitted in the range 0.2-1 TeV, for different EBL SEDs (as labelled). 
The dashed line corresponds to an EBL SED as low as the galaxy counts limit, and including 
galaxy evolution effects (according to \cite{prim01}).
The distance of the points over each curve represents visually the intrinsic hardness 
of the source spectra. 
Red: H.E.S.S. data. Blue: MAGIC data. Black: past results.
The sources are (from left to right): Mkn\,421, Mkn\,501, 1ES\,2344+514, Mkn\,180, 1ES\,1959+650,
PKS\,2005-489, PKS\,2155-304, 1ES\,1426+428, H\,2356-309, 1ES\,1218+304, 1ES\,1101-232,
PG\,1553+113 (details in \cite{matthias}, Table 1).
}
\label{all}       
\vspace*{-0.2cm}
\end{figure}

However, even if theoretically feasible, also this alternative seems highly unlikely.
If intrinsic, such pile-up features should be directly visible in the 
observed VHE spectra of closer, less absorbed HBLs like Mkn 421 and Mkn 501
(if $\Gamma_{\rm int}$ was $\approx0$, they should show $\Gamma_{\rm obs}\lesssim0.5$).
This is in contrast with observations, 
unless assuming a dependence of the source parameters on redshift 
such that the corresponding features always disappear due to EBL absorption.
It is difficult to justify such a fine-tuning on a relatively
small redshift range, even assuming an evolution of blazar properties.
Though the sample of objects is still too limited to settle this issue definitively, 
growing data seem to corroborate this conclusion. 
Figure 7 shows the observed photon indices of all TeV blazars detected so far
as a function of redshift, in comparison with  the expected index for a
$\Gamma=1.5$ intrinsic spectrum.
If the EBL density is high, the very strong dependence
on redshift would require a dramatic change of properties between 
TeV blazars at $z\sim0.2$ and $z\lesssim0.1$, with all 4 farther objects 
showing such hard features (which would thus seem rather common) 
and none of the 8 closer objects.
There seems to be no reason why the jet emission mechanisms  
in blazars should know the level of the EBL with such precision.
A low EBL level instead, close to the galaxy counts limit, does not impose 
such difference, making the TeV properties similar among different HBLs 
in this redshift range, 
as observed in all the other energy bands of the SEDs.

\section{Observational tests}
A straightforward step to test this conclusion 
is to look for independent evidence that HBLs
can indeed have much harder TeV spectra than assumed.
This can be done focussing in particular 
on lower ($z<0.1$) and higher ($z>0.3$) redshift objects.
At low redshift absorption is less, so hard features can be directly
measured in the observed TeV spectra (e.g. $\Gamma_{obs}\lesssim1$).
New detections and a monitoring program will increase the chances to 
catch an object in these special states, should they occur seldom.
\begin{figure}
\centering
\includegraphics*[width=0.26\textwidth]{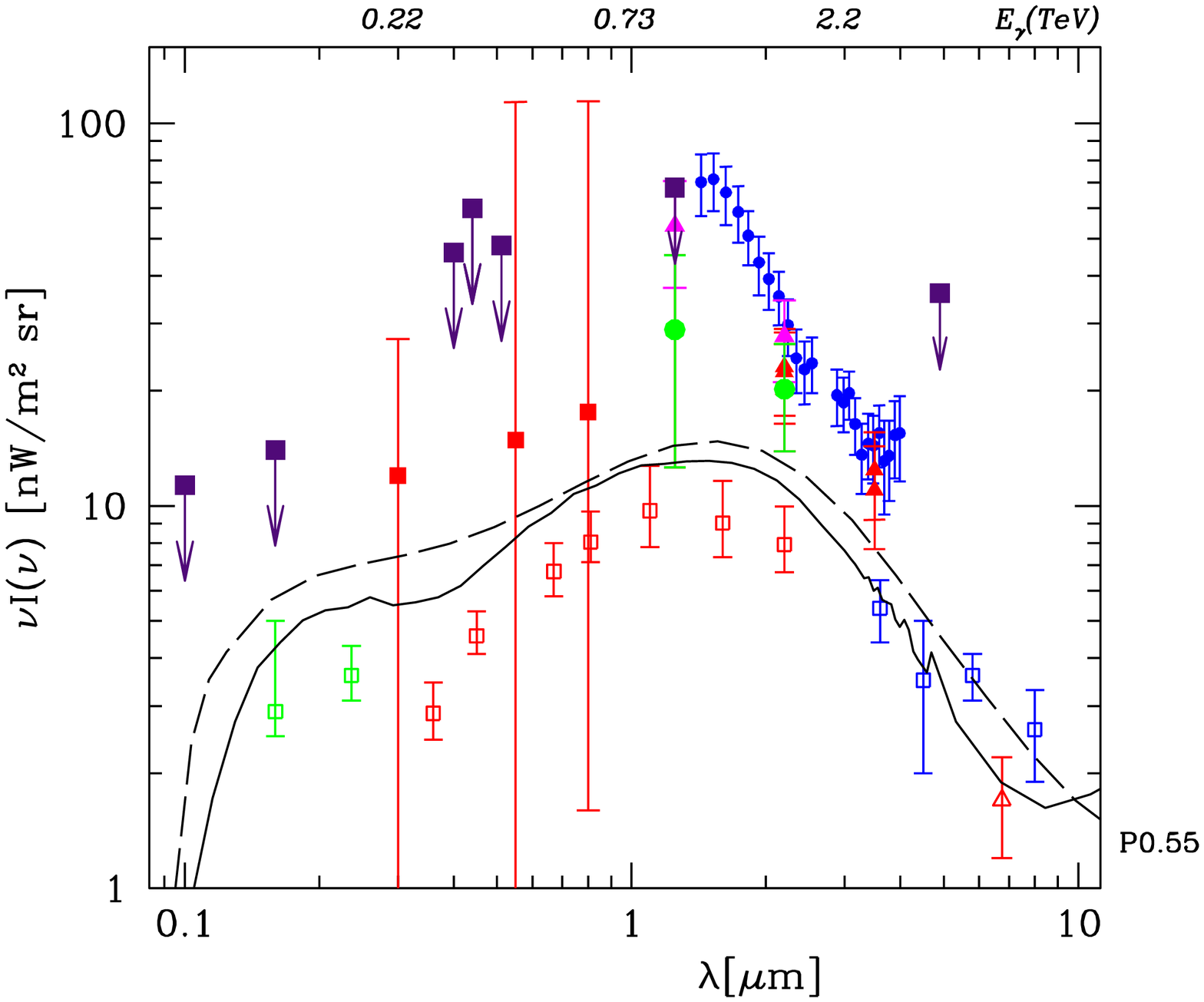}
\hspace{-0.2cm} \includegraphics*[width=0.21\textwidth]{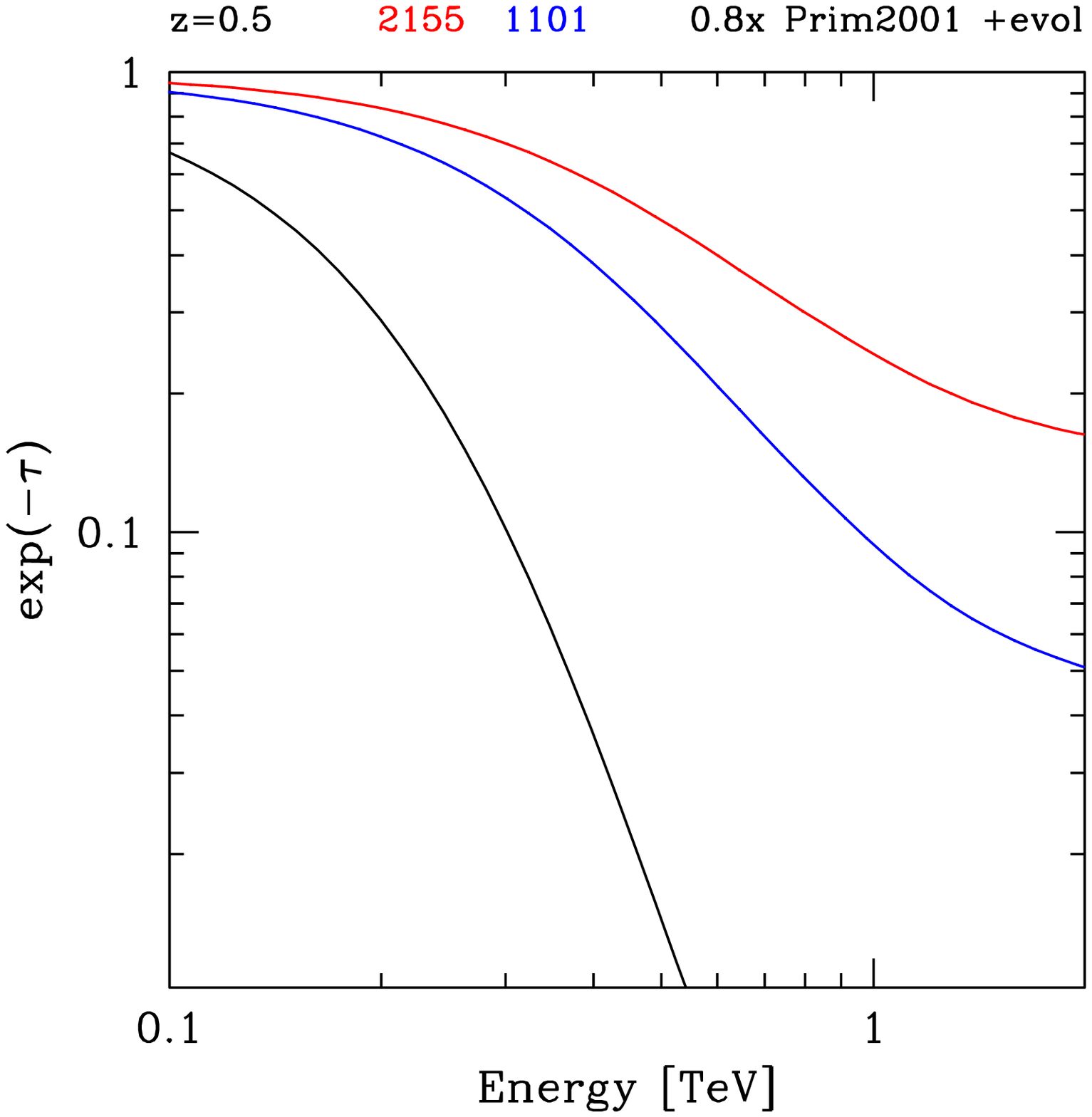}
\caption{Left, full line: Primack 2001 model\cite{prim01},
used to calculate the absorption properly including galaxy evolution effects.
The original model was scaled down $0.8\times$ to match the new EBL upper limit
(dashed line). 
Right: corresponding attenuation factor for a source at z=0.5 (black),
z=0.116 (red) and z=0.186 (blue).}
\label{highz}       
\vspace{-0.1cm} 
\end{figure}
Observations of high-redshift targets, 
in addition to extending the coverage of the $\Gamma-z$ plane (Fig. 7),
could reveal objects whose reconstructed spectrum is still very hard  
even with an EBL flux as low as the galaxy counts limit.
This would provide evidence for intrinsically hard spectra,
though in such case the possibility of exotic scenarios
invoking a modification of the physics of the $\gamma$-$\gamma$ interaction
would not be formally excluded.
With a low EBL level, objects at z=0.4-0.5 
should be well within the possibilities of the current CT generation.
Figure 8 shows the optical depth expected for a source at z=0.5
using the EBL model by \cite{prim01}, when 
scaled down to match the EBL upper limits.
This model was adopted to properly include galaxy evolution effects, which become 
important at larger redshifts.
Though the spectrum is expected very steep, up to 300 GeV
the attenuation is not dramatically higher than for the already detected 
TeV sources, and only a factor $\sim$2 around 150 GeV. 
Sources at such distances therefore could be detectable by current CT also outside 
exceptional (and rare) high-flux states.

The best targets for such observations are HBLs with 
the appropriate declination to be observable by CT at zenith
(to take advantage of the lowest possible energy threshold), and with
high X-ray brightness (indicating a large number of TeV electrons).
Using the BLLac samples and  criteria described in \cite{costagg},
a group of 3 objects best matching these requirements
have been found in the northern hemisphere (Fig. 9), suited for MAGIC and VERITAS.
They are therefore suggested for observations (Table 1).
A similar program for the southern hemisphere has been proposed for \hess.
%
\begin{table}[t]
\caption{High-z TeV candidates for MAGIC/VERITAS}
\centering
\label{tab}       
\begin{tabular}{lll}
\hline\noalign{\smallskip}
Name &  z & Sample  \\[3pt]
\tableheadseprule\noalign{\smallskip}
1ES 1106+244  & 0.460 &  Einstein Slew  \\ 
RX J1326.2+2933 & 0.431 & RASS \\ 
1ES 0219+428  & 0.444 &  Einstein Slew  \\ 
\noalign{\smallskip}\hline
\end{tabular}
\vspace*{-0.3cm}
\end{table}
\begin{figure}
\centering
  \includegraphics*[width=0.4\textwidth]{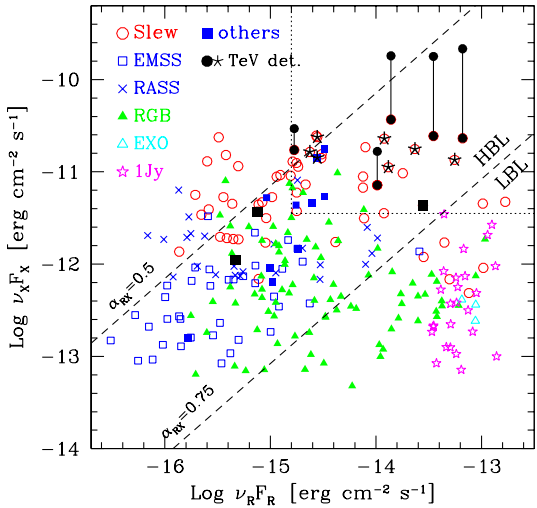}
\caption{BL Lac objects in the radio (5 GHz) and X--ray (1 keV)
$\nu F(\nu)$ plane (details in \cite{costagg}). Sources belonging to different samples have
different symbols, as labeled. Black crosses: new TeV detections since 
the proposal of the X-ray/radio Flux selection criterium (dotted rectangle).
Black squares: location of the proposed high-z sources for MAGIC/VERITAS
(Table 1). If at redshift $\sim0.1$, their location would be in the upper 
part of the rectangle (a change of redshift moves an object
approximately along the lines of constant $\alpha_{\rm RX}$).
}
\label{figf}       
\vspace*{-0.2cm}
\end{figure}
\section{Conclusions}
Showing an unexpected hardness for their redshift,
the combined \hess spectra of the HBLs 1ES\,1101-232 and H\,2356-309
(now supported by two further TeV sources) provide
strong circumstantial evidence for a low density of the EBL.
With minimal assumptions on the blazar properties, 
the \hess spectra allow the derivation of an upper limit on the EBL in the Opt-NIR
range which is very close to the lower limit given by the integrated 
light of resolved galaxies.
This result indicates that the EBL at these wavelengths is strongly dominated
by the direct starlight from galaxies, and thus excludes 
a large contribution from other sources, in particular Pop III stars.
At the same time, this result strongly reduces the uncertainty on the reconstruction 
of the blazar intrinsic spectra (up to few TeV), thus allowing 
a more accurate modelling and understanding of the overall SED.
However, additional detections and monitoring by the new-generation CTs
are needed to better sample the blazars' behaviour
at VHE, to further test this conclusion and possibly 
to pin down the diffuse background also above a few microns.

\vspace*{-0.4cm}
\begin{acknowledgements}
The support of the Namibian authorities and of the University of Namibia
in facilitating the construction and operation of H.E.S.S. is gratefully
acknowledged, as is the support by the German Ministry for Education and
Research (BMBF), the Max Planck Society, the French Ministry for Research,
the CNRS-IN2P3 and the Astroparticle Interdisciplinary Programme of the
CNRS, the U.K. Particle Physics and Astronomy Research Council \\ 
(PPARC),
the IPNP of the Charles University, the South African Department of
Science and Technology and National Research Foundation, and by the
University of Namibia. We appreciate the excellent work of the technical
support staff in Berlin, Durham, Hamburg, Heidelberg, Palaiseau, Paris,
Saclay, and in Namibia in the construction and operation of the
equipment.
This research has made use of the NASA/IPAC Extragalactic Database (NED) 
which is operated by the Jet Propulsion Laboratory, California Institute of Technology,
under contract with the National Aeronautics and Space Administration.
\end{acknowledgements}


\vspace*{-0.4cm}
\begin{thebibliography}{3}
\vspace*{-0.2cm}
\bibitem{hauser}Hauser, M.G \& Dwek, E.: The Cosmic Infrared Background: measurements and implications.
ARAA, {\bf 39}, 249 (2001)
\bibitem{santos}Santos, M.R. et al.: The contribution of the first stars to the cosmic
infrared background. MNRAS, {\bf 336}, 1082-1092 (2002)
\bibitem{madausilk}Madau, P. \& Silk, J.,: Population III and the near-infrared background 
excess. MNRAS, {\bf 359L}, 37-41 (2005)
\bibitem{icrc}Aharonian, F.A.: TeV blazars and the Cosmic Infrared Background radiation. 
{\it Invited, Rapporteur, and Highlight Papers. Proc. 27th ICRC (Hamburg)} (ed Schlickeiser R.), 
p.250-262 (2001), astro-ph/0112314 
%
\bibitem{nature}Aharonian, F.A., et al. (H.E.S.S. collab.): A low level of extragalactic background light as revealed 
by $\gamma$-rays from blazars. Nature, {\bf 440}, 1018-1022 (2006)
\bibitem{matthias}
Beilicke, M. for the \hess collab., these proceedings.
%
\bibitem{costa}
Costamante, L. et al.: Constraining the cosmic background light with four BL Lac TeV spectra.
NewAR, {\bf 48}, 469-472 (2004).
%
\bibitem{hegra}Aharonian, F.A. {\it et al.} Observations of H1426+428 with HEGRA. 
Observations in 2002 and reanalysis of 1999\&2000 data. 
A\&A, {\bf 403}, 523-528 (2003)  
\bibitem{hess2155}Aharonian, F.A., et al. (H.E.S.S. collab.): H.E.S.S. observations of PKS 2155-304. 
A\&A, {\bf 430}, 865-875 (2005)
\bibitem{dwek} Dwek, E. et al: The Near-Infrared Background: Interplanetary Dust or Primordial Stars?
ApJ, 635, 784-794 (2005)
\bibitem{prim01}Primack, J.R., et al.:
Probing Galaxy Formation with High-Energy Gamma Rays. AIP Conf. Proc., {\bf 558}, 463-478 (2001)
\bibitem{gg98}Ghisellini, G. et al.:
A theoretical unifying scheme for $\gamma$-ray bright blazars. MNRAS {\bf 301}, 451-468 (1998)
\bibitem{padovani}
Padovani, P.,  these proceedings.
\bibitem{malkov}Malkov, M.A. \& O'C Drury, L., Nonlinear theory of diffusive acceleration 
of particles by shock waves.  {\it Rep. Prog. Phys.}, {\bf 64}, 429-481 (2001)
%
\bibitem{schick2} Schlickeiser, R.: Cosmic ray astrophysics. Springer-Verlag, XV+519pp (2002)
%
\bibitem{vainio} Vainio, R.: Diffusive Shock Acceleration. 
Proc. Conf. ``Plasma turbulence and energetic particles in astrophysics", 
Cracow (Poland), p.232-245, (1999)
%
\bibitem{bernstein}Bernstein, R.A. et al.: The First Detections of the Extragalactic Background 
Light at 3000, 5500, and 8000 \AA. I. Results. ApJ, {\bf 571}, 56-84 (2002)
%
%
\bibitem{hs} Sauge`, L. \& Henri, G.: 
TeV Blazar $\gamma$-ray Emission Produced by a Cooling Pileup Particle Energy Distribution Function.
ApJ, {\bf 616}, 136-146 (2004)
\bibitem{pp} Park, B.T. \& Petrosian,V.: Fokker-Planck Equations of Stochastic Acceleration: A Study 
of Numerical Methods. ApJS {\bf 103} 255-267 (1996)
\bibitem{katar} Katarzynski, K. et al.: Hard TeV spectra of blazars and the constraints to the 
infrared intergalactic background. MNRAS, {\bf 368}, L52-L56 (2006)
\bibitem{costagg} 
Costamante, L. \& Ghisellini, G.: TeV candidate BL Lac objects. A\&A {\bf 384}, 56-71 (2002)
\bibitem{elbaz}Elbaz, D. {\it et al.}  The bulk of the cosmic infrared background resolved by ISOCAM. A\&A, {\bf 384}, 848-865 (2002) 
\bibitem{dole}Dole, H. et al.: The cosmic infrared background resolved by Spitzer. Contributions of mid-infrared galaxies to the far-infrared background. A\&A, {\bf 451}, 417-429 (2002) 
\bibitem{mattila}Mattila, K.: Has the Optical Extragalactic Background Light Been Detected?  ApJ, {\bf 591}, 119-124 (2003) 
\bibitem{wolter}Wolter, A., et al.: X-ray variability and prediction of TeV emission 
in the HBL 1ES\,1101-232. A\&A, {\bf 357}, 429-436 (2000) 
\end{thebibliography}
\end{document}